\documentclass[journal]{IEEEtran}

\ifCLASSINFOpdf
\else
\fi

 \usepackage{graphicx}

\usepackage{flushend}

\begin{document}
%
\title{Design and Analysis of a Multi-Carrier  Differential Chaos Shift Keying Communication System}
%
%
%

\author{Georges Kaddoum$^{*}$, Fran\c cois-Dominique Richardson,  Fran\c cois Gagnon

\thanks{G. Kaddoum, F.-D. Richardson and F. Gagnon are with University of Qu\'{e}bec, ETS,
LaCIME Laboratory, 1100 Notre-Dame west, H3C 1K3, Montreal, Canada (e-mail:
georges.kaddoum@lacime.etsmtl.ca; francois.richardson@lacime.etsmtl.ca; francois.gagnon@etsmtl.ca)}
}


\maketitle

{\let\thefootnote\relax\footnotetext{* This work has been supported in part by Ultra Electronics TCS and the Natural Science and Engineering Council of Canada as part of the 'High Performance Emergency and Tactical Wireless Communication Chair' at \'{E}cole de technologie sup\'{e}rieure.}}

\begin{abstract}
A new Multi-Carrier Differential Chaos Shift Keying (MC-DCSK) modulation is presented in this paper. The system endeavors to provide a good trade-off between robustness, energy efficiency and high data rate, while still being simple compared to conventional multi-carrier spread spectrum systems. This system can be seen as a parallel extension of the DCSK modulation where one chaotic reference sequence is transmitted over a predefined subcarrier frequency. Multiple modulated data streams are transmitted over the remaining subcarriers. This transmitter structure  increases the spectral efficiency of the conventional DCSK system and uses less energy. The receiver design makes this system easy to implement where no radio frequency (RF) delay circuit is needed to demodulate received data. Various system design parameters are discussed throughout the paper, including the number of subcarriers, the spreading factor, and the transmitted energy. Once the design is explained, the bit error rate performance of the MC-DCSK system is computed and compared to the conventional DCSK system under multipath Rayleigh fading  and an additive white Gaussian noise (AWGN) channels. Simulation results confirm the advantages of this new hybrid design. 
\end{abstract}




%
\IEEEpeerreviewmaketitle

\section{Introduction}
\IEEEPARstart{A}{s} computing devices become ubiquitous, a plurality of challenges emerge from the various communications paradigms. Some researchers have envisioned that there will be ``Seven Trillion Wireless Devices Serving Seven Billion People by 2020''~\cite{David2010}. In this perspective, spectral and power efficiency, interference resistance, security and channel agnosticism are, and will continue, to be top requirements for wireless communication systems. 

Mobile wireless communications performances are deteriorated by device hardware and the propagation environments~\cite{Mingo2004}. Fading channels, for instance in Vehicle-to-Vehicle communication where moving scatterers cause a non-wide-sense stationary uncorrelated scattering (WSSUS) behavior~\cite{Karedal2009}, are typically harsh environments for mobile communications. In order to get optimal communication systems in varying channels, many techniques can be employed. One is the use of multi-carrier systems, such as OFDM, that have a high resilience to selective channels if the bandwidth of each subcarrier is smaller than the coherence bandwidth; non-coherent communication systems make up the other. It has been stated that non-coherent systems can outperform coherent ones in fast frequency and time-varying channels, mainly because of the spectral inefficiency inherent to the insertion of pilots~\cite{LeSaux2005}.

Several combinations of multi-carrier transmission and Code Division Multiple Access (CDMA), like Multi-Carrier CDMA (MC-CDMA), Multi-Carrier Direct-Sequence CDMA (MC-DS-CDMA) and Orthogonal Frequency Code Division Multiplexing (OFCDM) are proposed in the literature~\cite{Hanzo03, Nee00, kond96}. In MC-CDMA, one-bit chips are spread over $M$ subcarriers in the frequency domain~\cite{Hanzo03}, while for MC-DS-CDMA, time and frequency spreading is used (\textit{i.e.} TF-domain spreading)~\cite{kond96}. Time-domain spreading is employed to increase the processing gain in each subcarrier signal, while frequency domain spreading is used to increase the total processing gain.

The chaotic signal has a "sensitive dependence upon initial conditions" property that allows the generation of a theoretical infinite number of uncorrelated signals. Those signals have been shown to be well suited for spread-spectrum modulation because of their inherent wideband characteristic~\cite{Lau03} \cite{Kur05} \cite{Val12}. Chaotic modulations thus have similar advantages as other spread-spectrum modulations, \textit{exempli gratia}, including the mitigation of fading channels and jamming resistance. The low probability of interception (LPI) performance of chaotic signals~\cite{Yu2005} agrees with military scenarios and in densely populated environments \cite{Lyn10}. In addition, chaos-based sequences give good results in comparison to Gold and independent and identically distributed sequences for reducing the peak-to-average power ratio (PAPR)~\cite{Vitali2006}.


A proposed system with a non-coherent receiver, named a differential chaos shift keying (DCSK) system, in which chaotic synchronization is not used on the receiver side to generate an exact replica of chaotic sequence but only requires frame or symbol rate sampling \cite{6328505}. This system delivers a good performance in multipath channels~\cite{Kol98a} \cite{Xia04}. Furthermore, differential non-coherent systems are better suited than coherent ones for time and frequency selective channels~\cite{LeSaux2005}. In the DCSK system, each bit duration is divided into two equal slots. In the first slot, a reference chaotic signal is sent. Depending on the bit being sent, the reference signal is either repeated or multiplied by the factor $-1$ and transmitted in the second slot. A significant drawback of DCSK is that half the bit duration is spent sending non-information-bearing reference samples~\cite{Lau03}. This can be accounted as being energy-inefficient and a serious data rate reducer.  The analytical performance derivation of DCSK communication system is studied in \cite{kad12wpc}  for fading channels  and in \cite{Fan13} \cite{Fan13i} \cite{Wei11} \cite{Jing10} for cooperative schemes. The transmission security of DCSK system is improved in \cite{KadFDR2012a}. In \cite{Hua12}, the spectral efficiency of the DCSK is improved, but the system receiver requires an RF delay line, which is not easy to implement because of the wide bandwidth involved. In a study to overcome the problem of RF delay in DCSK systems, Xu \textit{et al.} proposed a Code Shifted Differential Chaos Shift Keying (CS-DCSK) system~\cite{Xu11}. In their system, the reference and the information bearing signals are separated by Walsh code sequences, and then transmitted in the same time slot. For such systems, there is no need for a delay line at the receiver end. An improved version of the high spectral efficiency DCSK system by \cite{Xu11} is presented in \cite{kad12}, where chaotic codes are used instead of Walsh codes, with different receiver structures. Because of these advantages, some ultrawideband systems based on DCSK or Frequency Modulation DCSK (FM-DCSK) modulations have been proposed for wireless personal area networks  \cite{Cima07, chi08, Min10}.

In this paper, we first introduce a new design of a multi-carrier DCSK system (MC-DCSK). The system is a hybrid of multi-carrier and DCSK modulations. On the transmitter side, one of the $M$ subcarriers is assigned to transmitting the reference slot, while the other frequencies will carry the data slots. In this case, just one chaotic reference is used to transmit $M-1$ bits, which saves the transmitted bit energy and  increases the data rate. After the subcarriers are removed, a parallel demodulation is achieved to quickly recover the transmitted bits. The proposed system solves the RF delay line problem mentioned in~\cite{Xu11}, provides from the properties of DCSK  system in terms of resistance to interference, increases the data rate, and optimizes the transmitted energy of the DCSK system with a simple transmitter/receiver design compared to conventional multi-carrier spread spectrum systems (i.e MC-DS-CDMA).

Secondly, we thoroughly analyze the performance under multipath Rayleigh fading and AWGN channels, without neglecting the dynamic properties of chaotic sequences. In our computation approach, the transmitted bit energy is not considered as constant. Many approaches have been considered for computing the bit error rate performances of the DCSK system, such as Gaussian approximation (GA) \cite{Kol98a} \cite{Xia04}. This approximation assumes that the correlator output follows the normal distribution. Applied to the DCSK system over an AWGN or multipath channel in \cite{Zho08}, this method provides rather good estimates of the bit error rate (BER) for very large spreading factors, but when the spreading factor is small, the results produced by the Gaussian approximation method are rather disappointing. 
Another accurate computation methodology  is developed in  \cite{Law03}, \cite{Law06} and \cite{Zho09} to compute the BER performance DCSK over different wireless channels. Their approach enables the dynamic properties of the chaotic sequence by integrating the BER expression for a given chaotic map over all possible chaotic sequences for a given spreading factor. This latter method is compared to the BER computation under the Gaussian assumption in \cite{Zho09}, and seems more realistic to match the exact BER. However, as indicated in \cite{Zho09, Law06}, the drawback of the proposed method is the high calculation difficulty. 
Since approaches that have been previously presented are either invalid for small spreading factors or involving highly complex computations, we extend a simple and accurate method in this paper for computing the exact performance for a single-user MC-DCSK system for low spreading factor. The system is evaluated first over an AWGN to highlight the problem of non constant bit energy. Otherwise, for high spreading factor the GA is assumed. Then, the performance of MC-DCSK system is evaluated under a multipath Rayleigh fading channel. The proposed method includes the computation of the probability density function of the chaotic bit energy  for low spreading factor and the integration of the BER over all possible values of the PDF. The advantage of this method lies in the fact that it gives an exact BER expression without neglecting the dynamical properties of chaotic sequences with low computing charges.

Thirdly, we derive the analytical bit error rate expressions, and we show the accuracy of our analysis by matching the  numerical performance. We can conclude that the proposed system can be suited for Wireless Sensor Network (WSN) applications ~\cite{Nayak2010}, which are power-limited and evolve in harsh environments and high resistance to multipath interference.  

The remainder of this paper is organized as follows. In section~\ref{sec:Chaos}, the characteristics of chaos-based systems are described with an emphasis on DCSK. The third section covers the architecture of the MC-DCSK system. The energy and spectral efficiencies of the system are examined in section~\ref{EnergyEfficiency}. The performance analysis is explained in section~\ref{sec:PerformancesAnalysis}. Simulation results and discussions are presented in section~\ref{sec:Simulation}, and concluding remarks are presented in section~\ref{sec:Conclusion}.

\section{DCSK communication system\\ and weakness points}
\label{sec:Chaos}
In this section, the DCSK communication system, together with its weakness points, are discussed. The conventional DCSK modulation scheme will be used in section~\ref{sec:Simulation} as a comparative to illustrate the performance enhancements obtained from the main contribution of this paper.  

\subsection{DCSK communication system}
\begin{figure}[t]
\centering 
\includegraphics[width=8 cm]{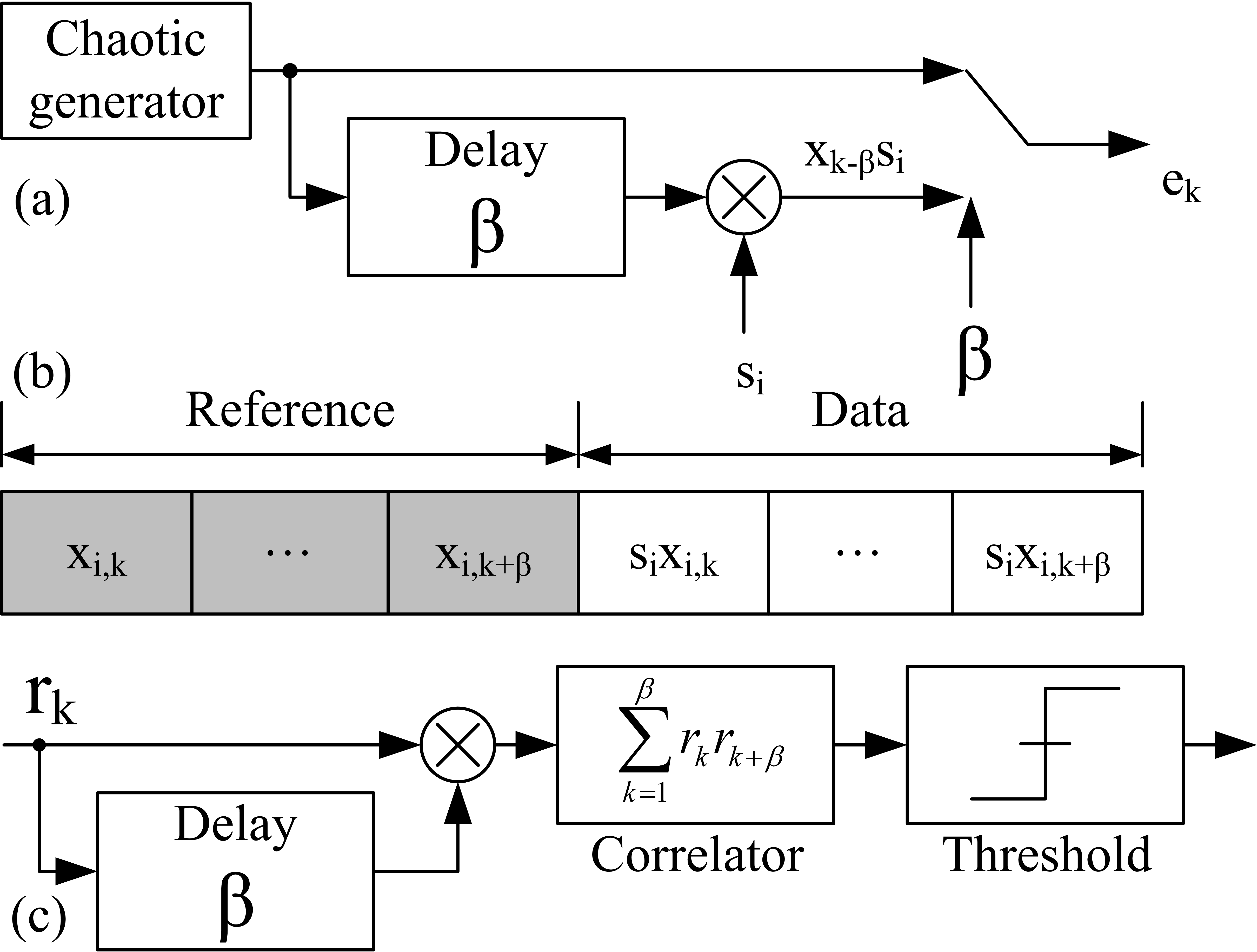}
\caption {Block diagram of the general structure of the DCSK communication system. (a) is the DCSK transmitter, (b) represents the DCSK frame (c) is the DCSK receiver. \label{DCSK}} 

\end{figure}

 As shown in Fig.~\ref{DCSK}, within the modulator, each bit $s_i=\lbrace -1, \: +1 \rbrace$ is represented by two sets of chaotic signal samples, with the first set representing the reference, and the second carrying data. If $+1$ is transmitted, the data-bearing sequence is equal to the reference sequence, and if $-1$ is transmitted, an inverted version of the reference sequence is used as the data-bearing sequence. 
Let $2\beta$ be the spreading factor in DCSK system, defined as the number of chaotic samples sent for each bit, where $\beta$ is an integer. 
During the $i^{th}$ bit duration, the output of the transmitter $e_{i,k}$ is
\begin{equation}\label{et}
e_{i,k} = \left\{ \begin{array}{l}
x_{i,k} \,\,\,\,\,\,\,\,\,\,\,\,\,\,\,\,\,\,\,{\rm for }\,\,1 < k \le \beta ,\\ 
s_i x_{i,k - \beta \,\,\,} \,\,  \,\,{\rm for }\,\,\beta < k \le 2\beta , \\ 
\end{array} \right.
\end{equation}
\\
where $x_k$ is the chaotic sequence used as reference and $x_{k-\beta}$ is the delayed version of the reference sequence.
 
Fig.~\ref{DCSK} illustrates that the received signal $r_{k}$ is correlated to a delayed version of the received signal $r_{k + \beta }$ and summed over a bit duration $T_b$ (where $T_b=\beta T_c$ and $T_c$ is the chip time) to demodulate the transmitted bits. The received bits are estimated by computing the sign of the output of the correlator (i.e., see Fig.~\ref{DCSK} (c) the DCSK receiver). 


\subsection{Weakness of DCSK }
 
In this paper, we are not working on improving the inherent lack of security of non-coherent systems, as the security issue was addressed in our previous work~\cite{KadFDR2012a}, where a secure chaos-based multi-carrier communication system was proposed. Our focus in this work is on the spectral and energy efficiencies having a good performance. 

As shown in Fig.~\ref{DCSK}, half the bit duration time is spent sending a non-information-bearing reference. Therefore, the data rate of this architecture is seriously reduced compared to other systems using the same bandwidth, leading to a loss of energy. The reference sequence dissipates half the energy of each bit.

\section{Multi-carrier DCSK system architecture}
\label{sec:SystemArchitecture}
\begin{figure*}[t!]
\centering
\includegraphics[width=18 cm]{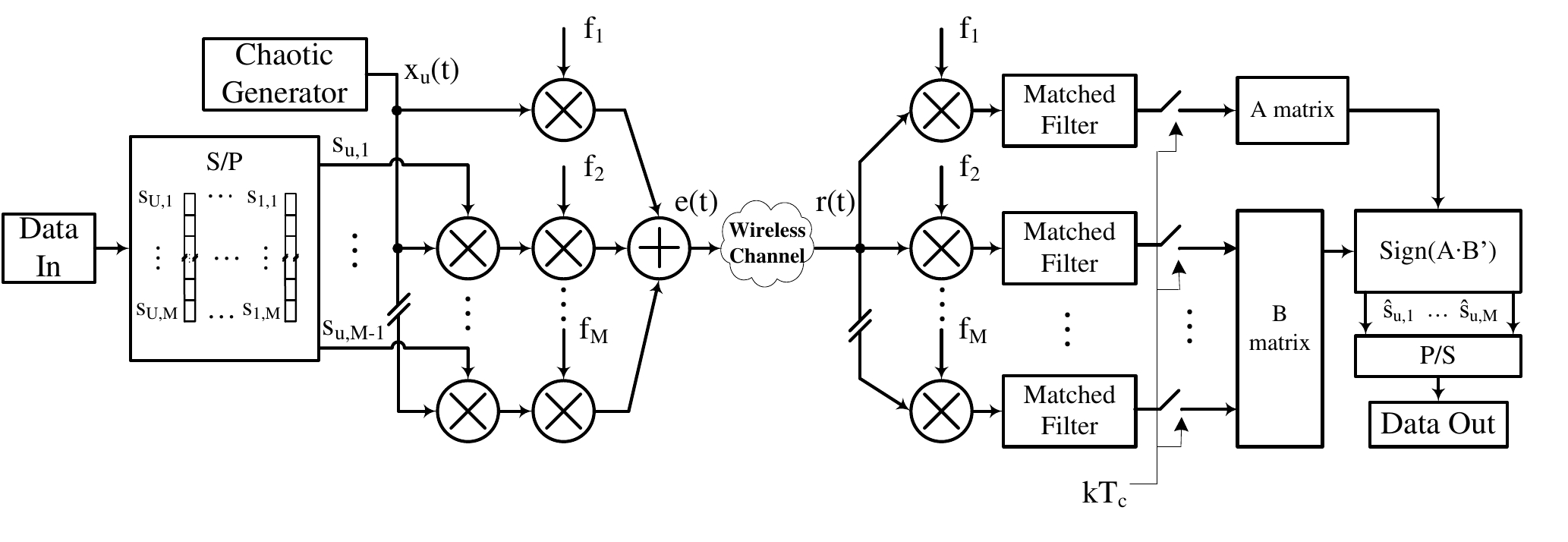}
\caption{Block diagram of the MC-DCSK system \label{fig:TxRx}}

\end{figure*}

The system's architecture is intended to be of low complexity. Numerous extensions could be performed to this system for different performance optimizations. The system presented here is in its most elementary form.

\subsection{Chaotic generator}
In this paper, a second-order Chebyshev polynomial function (CPF) is employed 

\begin{equation}\label{cpf}
 x_{k+1}=1-2x_{k}^{2} \cdot
\end{equation}

This map is chosen for the ease with which it generates chaotic sequences and the good performance \cite{Kad08cs}. In addition, chaotic sequences are normalized such that their mean values are all zero and their mean squared values are unity, i.e., $E(x_{k})=0$ and $E(x_{k}^{2})=1$.

\subsection{The transmitter}
\label{sec:Tx}
The MC-DCSK system benefits from the non-coherent advantages of DCSK and the spectral efficiency of multi-carrier modulation. 
In this system, the input information sequence is first converted into $U$ parallel data sequences ${\bf s}_{\bf u}$ for $u=1,\: 2,...U$.

\begin{equation}
{\bf s}_{\bf u}  = \left[ {s_{u,1} , \ldots ,s_{u,i} , \ldots ,s_{u,M - 1} } \right],
\end{equation} 
\\
where $s_{u,i}$ is the  $i^{th}$ bit of the $u^{th}$ sequence data and $M-1$ is the number of data per $u^{th}$ sequence.

As shown in Fig.~\ref{fig:TxRx}, a reference chaotic code $x_u(t)$ to be used as a reference signal and spreading code. 
After a serial-to-parallel conversion, the ${M-1}$ bits stream  of the $u^{th}$ data sequence are spread due to multiplication in time with the same chaotic spreading code $x_u(t)$. 
 \begin{equation}
 x_u (t) = \sum\limits_{k =  1  }^{ \beta } {x_{u,k} } h(t-kT_c),
 \end{equation}  
 \\
where, $\beta$ is the spreading factor, $h(t)$ is the square-root-raised-cosine filter. This filter is band-limited  and is normalized to have unit energy. Let $H(f)=\textit{F} 
\left\{ {h(t)} \right\}$, where $\textit{F} $ denotes a Fourier transform. It is assumed that $H(f)$ is limited to $[-B_c/2, \: B_c/2]$ which satisfies the Nyquist criterion with a rolloff factor $\alpha$ ($0 \le \alpha  \le 1 $) where $B_c=(1+\alpha)/T_c$.

The chaotic signal $x_u(t)$ modulates the first subcarrier as reference, after which the data signals spread  by ${M-1}$ modulate the ${M-1}$ subcarriers. 

Therefore, the transmitted signal of the MC-DCSK is given by:

\begin{equation}
\begin{array}{l}
 e(t) =  {x_{u} (t)\cos (2\pi f_1 t + \phi _1 )}  +  \\ 
 \sum\limits_{i = 2}^M {  {s_{u,i}^{} (t)x_{u} (t)\cos (2\pi f_i t + \phi _i )} }  \\ 
 \end{array} ,
\end{equation}
\\
where $\phi _i $ represents the phase angle introduced in the carrier modulation process. In this paper, we normalize the transmitted energy in every subcarrier.
 
For the MC-DCSK, the modulated subcarriers are orthogonal over the chip duration. Hence, the baseband frequency corresponding to the $i^{th}$ subcarrier is $f_{i}= f_{p} + i/T_c$, where $f_p$ is the fundamental subcarrier frequency. The minimum spacing between two adjacent subcarriers equals $\Delta=(1 +\alpha)/T_c$, which is a widely used assumption \cite{kond96}.

\begin{figure}[htb!]
\centering
\includegraphics[width=8.5 cm]{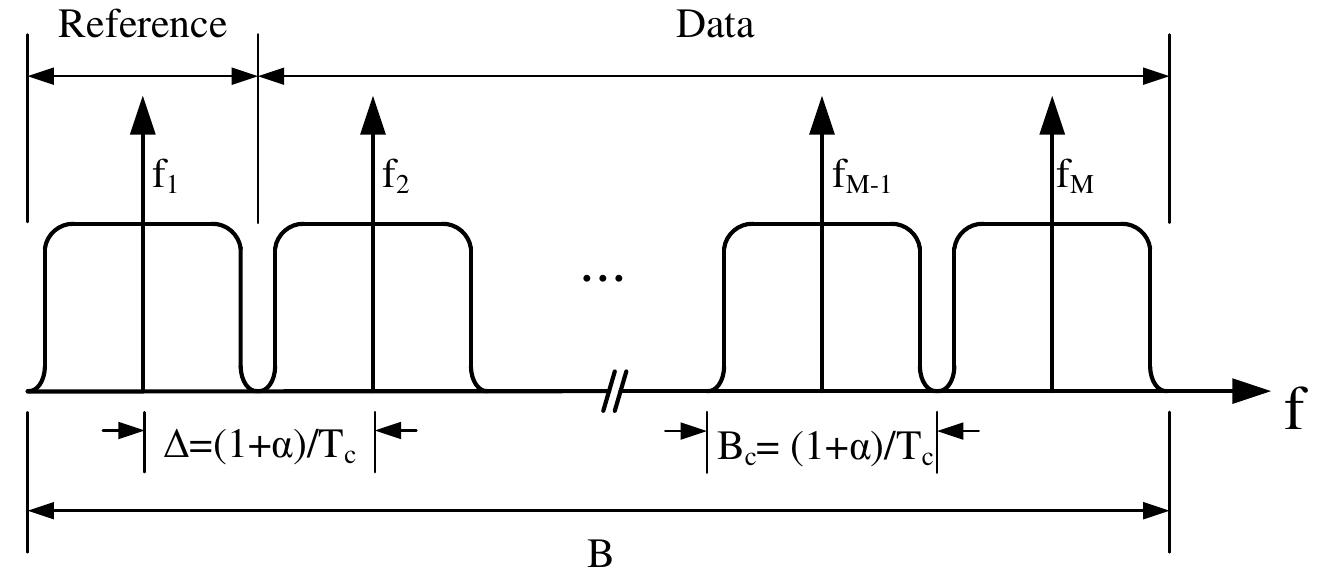}
\caption{The power spectral density of a band-limited MC-DCSK system. \label{psd_mc_DCSK}}

\end{figure}

Fig. \ref{psd_mc_DCSK} shows the power spectral density (PSD) of the MC-DCSK system. Let $B$ be the total bandwidth of the proposed system. When both bit duration $T_b$ and $B$ are set, the chip duration $T_c$ as well as the spreading factor $\beta$ depend on the number of subcarrier $M$, the bandwidth $B_c$ of each subchannel or the subcarrier spacing $\Delta$. In our design, we divide the total band $B$ into $M$ equi-width frequency bands, as shown in Fig. \ref{psd_mc_DCSK}, where all bands are disjoint. The bandwidth of each subcarrier band $B_c$ is:

\[B_c=(1+ \alpha)/T_c \cdot \]

The total required bandwidth $B$ is:
\[B=M B_c ,\]

\[ B= M(1+ \alpha)/T_c \cdot \] 

Thus, the spreading factor  function of the system parameters is: 

\[ \beta = T_b/T_c , \]

\begin{equation} \label{sr_fc}
\beta  = \frac{{T_b B}}{{M(1 + \alpha )}} \cdot
\end{equation}

Finally, the received signal is given by:

\begin{equation}
r(t)= \sum\limits_{l = 1}^L {\lambda _{l}(t- \tau_l)}*e(t)+n(t),
\end{equation}
\\
where $L$ is the number of path, $\lambda_l(t)$ and $\tau_l$ are the channel coefficient and the appropriate delay of the $l^{th}$ path respectively, $*$ is the convolution operator,  and $n(t)$ is an wideband AWGN  with zero mean and power spectral density of $N_0/2$.

\begin{figure}[htb!]
\centering
\includegraphics[width=9 cm]{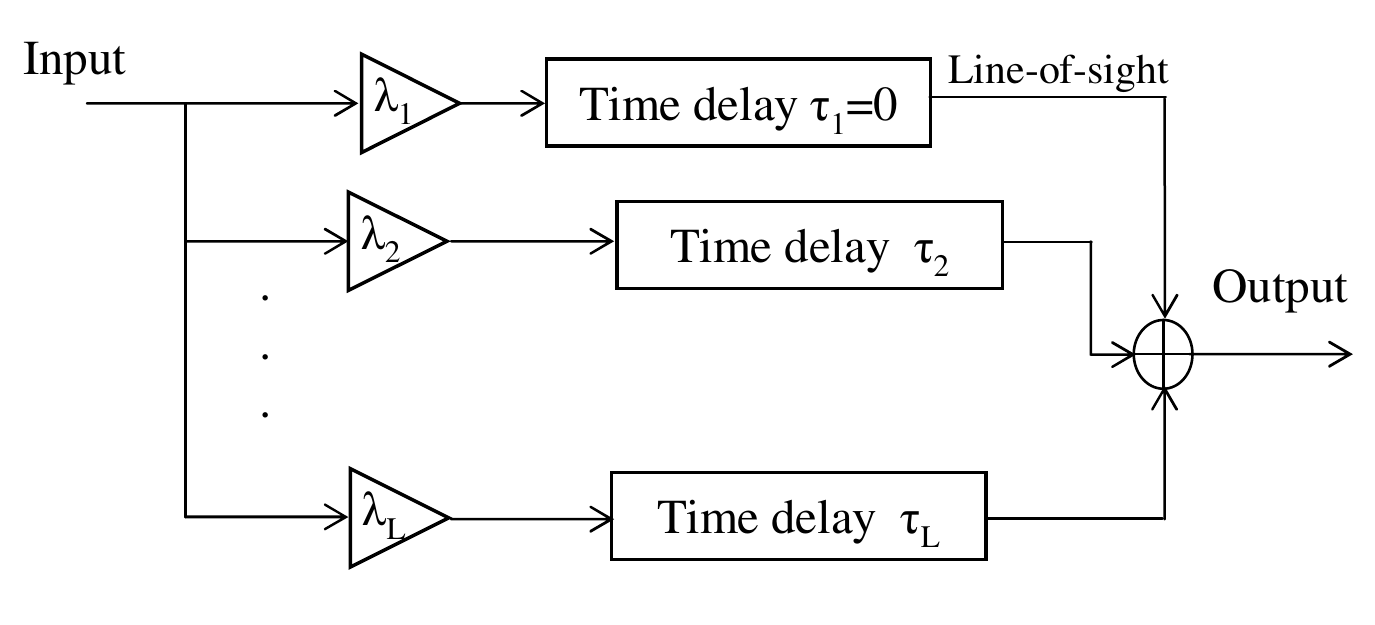}
\caption{Multipath Rayleigh fading model \label{chan_mod}}
\end{figure}

For our analysis, we choose a commonly used channel model in spread spectrum wireless communication systems  \cite{Rap96}, \cite{Xia04}, \cite{Chen13}. A two-ray Rayleigh channel model is used in \cite{Xia04}, \cite{Chen13}.  As  shown in Fig. \ref{chan_mod}, we consider a slow fading multipath channel with $L$ ($L \geq 2$) independent and Rayleigh distributed random variables. In this model,  $\lambda_{l}$ is the channel coefficient and $\tau_{l}$ is time delay of the $l^{th}$ path (i.e  for $L=1$  $\tau_{1}=0$ line-of sight). The Rayleigh probability density function of the channel coefficient is given by:

\begin{equation}
f_\lambda  (z) = \frac{z}{{\sigma ^2 }}e^{ - \frac{{z^2 }}{{2\sigma ^2 }}}  \: \:\: \:\: \:\: \: z>0  \cdot
\end{equation}
\\
where $\sigma >0$ is the scale parameter of the distribution.

In an AWGN case, the number of path is equal to one  $L=1$ with a unit channel coefficient $\lambda(t)=1$.

\subsection{The receiver}
\label{sec:Rx}

The block diagram of the MC-DCSK receiver is illustrated in Fig.~\ref{fig:TxRx}. One of the objectives of this design was to provide a robust receiver providing good performance. We consider a set of matched filters, each demodulating the desired signal of the corresponding carrier frequency $f_i$, and then the signals are sampled every $kT_c$ time. The outputs discrete signals are stored in matrix memory. The matrix implementation of the receiver simplifies the parallel data recovery, where the decoding algorithm is described as follow under AWGN channel: 

First, at the same time, the output of the first match is stored in matrix $P$ and the ${M-1}$ data signals are stored in the second matrix $S$, where:

\[P= (x_{u,1} + n_{u,1}, \: x_{u,2}+n_{u,2}, ... \: x_{u,\beta}+n_{u,\beta}) ,\] 
\\
where $n_{u,k}$ is the $k^{th}$  sample of additive Gaussian noise added to the reference signal.

The matrix $S$ is:

\[
S \hspace{-0.1 cm} = \hspace{-0.1cm} \left( \hspace{-0.2cm} {\begin{array}{*{20}c}
   {s_{u,1} x_{u,1} +n^{1}_{u,1} } &  \ldots  & {s_{u,1} x_{u,\beta } +n^{1}_{u,\beta}}  \\
    \vdots  &  \vdots  &  \vdots   \\
   {s_{u,M - 1} x_{u,1}  +n^{M-1}_{u,1} } &  \ldots  & {s_{u,M - 1} x_{u,\beta } +n^{M-1}_{u,\beta} }  \\ 
\end{array}} \hspace{-0.2cm} \right)\cdot
\]
\\
where $n^{i}_{u,k}$ is the $k^{th}$  sample of additive Gaussian noise added to the $i^{th}$ bit of $u^{th}$ data sequence.

Finally, after $\beta$ clock cycles, all the samples are stored, and the decoding step is activated.
The transmitted ${M-1}$ bits are recovered in parallel by computing the $\textrm{sign}$ of  the resultant vector of the matrix product:

\begin{equation}
\hat s_u  = \textrm{sign}(P \times S')\cdot
\end{equation}
\\
where $ \times $ is the matrix product and $'$ is the matrix transpose operator. In fact, this matrix product can be seen as a set of a parallel correlator where the reference signal multiplies each data slot, and the result is summed over the duration $\beta Tc$.


\section{Energy  efficiency }
\label{EnergyEfficiency}

The energy efficiency of the proposed system is improved as compared to the DCSK system. In fact, for the DCSK system, a new chaotic reference is generated for every transmitted bit, and in our case, one reference is shared with ${M-1}$ modulated bits. For a conventional DCSK system, the transmitted bit energy $E_b$ is:

\begin{equation}
E_b=E_{data}+E_{ref} \cdot
\end{equation}
\\
where $E_{data}$  and $E_{ref}$ are the energies to transmit the data and reference respectively. 
Without loss of generality, the data and the reference energies are equal:

\begin{equation}
  E_{data}=E_{ref}=T_c \sum\limits_{k = 1}^\beta  {x_{k}^2 } \cdot
 \end{equation} 
  
Then for DCSK system, the   transmitted  energy $E_b$ for a given bit $i$ is:

\begin{equation}
  E_b=2T_c \sum\limits_{k = 1}^\beta  {x_{k}^2 } \cdot
\end{equation} 

In the MC-DCSK system, one reference energy $E_{ref}$ is shared with $M-1$ transmitted bit, then the energy of one given bit is the sum of its data carrier energy and a part of the reference energy:

\begin{equation}
E_b=E_{data}+ \frac{{E_{ref}}}{{M-1}} \cdot
\end{equation}

In our system, the energies on the $M$ subcarriers are equal:

\begin{equation}
E_{data}=E_{ref}=T_c \sum\limits_{k = 1}^\beta  {x_{k}^2 } \cdot
 \end{equation}
 
The bit energy expression function of $E_{data}$ is:

\begin{equation}
E_b=\frac{{M }}{{M-1}}E_{data} \cdot
\end{equation}

To study the energy efficiency, we compute the transmitted Data-energy-to-Bit-energy Ratio (DBR):
\begin{equation}
\label{PowerRatio}
DBR=\frac{E_{data}}{E_{b}} \cdot
\end{equation}

For the MC-DCSK system the DBR is:

\begin{equation}
\label{PowerRatio}
DBR=\frac{M-1}{M} \cdot
\end{equation}

In a conventional DCSK system (i.e for $M=2$), half the energy  $E_b$ is transmitted into the reference  for each bit, and then the DBR is: 

\begin{equation}
\label{PowerRatio}
DBR=\frac{1}{2} \cdot
\end{equation}

\begin{figure}[t]
\centering
\includegraphics[width=9 cm]{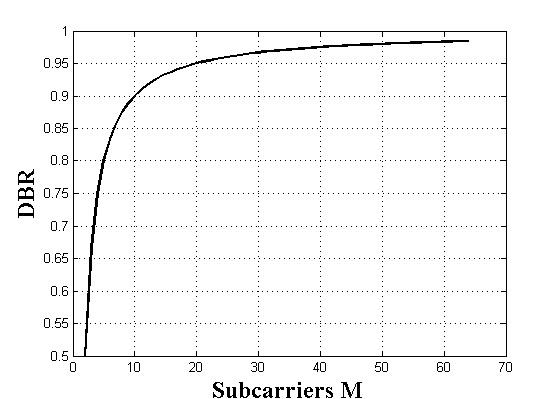}
\caption{DBR for a system for various amount of data subcarriers. \label{fig:PowerRatio}}

\end{figure}

As shown in Fig. \ref{fig:PowerRatio}, for $M=2$ where we have one new reference for every bit  bit, in this case, the MC-DCSK system is equivalent to a DCSK system with $DBR=\frac{1}{2}$. This means that $50\%$ of the bit energy $E_b$ is used  to transmit the reference used for one bit. For the same bit energy $E_b$, in MC-DCSK system, we can see for example that for $M > 20$, the reference energy accounts for less than $5\%$ of the total bit energy $E_b$ for each bit of the $M-1$ data stream. This mean that the energy used to transmits the reference is shared with $M-1$ bits.

\section{Performance analysis of MC-DCSK}
\label{sec:PerformancesAnalysis}

In this section, the performance of the MC-DCSK system is evaluated, and the analytical BER expression is derived under AWGN  and multipath Rayleigh fading channels.

\subsection{Derivation of  the BER expression}

To derive the analytical BER expression, the mean and the variance of the observation signal $D_{u,i}$ must be evaluated. With this aim, we start by mentioning some properties of chaotic signals which will be used later to analyse the statistical properties the observation signal. A chaotic generator is very sensitive to initial conditions, and we can deduce that the different chaotic sequences generated from different initial conditions are independent from each other. In addition, the independence between the chaotic sequence and the Gaussian noise is also true~\cite{Lau03}. For the normalized chaotic map with zero mean, the variance ($Var(.)$) is equal to one $(Var(x)= E(x^2)=1)$.


In our analysis, we assume that the largest multipath time delay is shorter than the bit duration:

\begin{equation}
0 <\tau_{L} << \beta 
\end{equation}

In this case, the intersymbol interference  (ISI) is negligible compared with the interference within each symbol due to multipath delay. However, when  $\tau_{L}$ increases, ISI increases and deteriorates the BER. In most practical applications, the condition $\tau_{L} << \beta$ holds, making our assumption justifiable \cite{Kol00}. Notwithstanding, we approve in the next section that the negligible ISI hypothesis when  $\tau_{L} << \beta$ is true, and we also show the limit of this negligible ISI assumption.  Moreover,  it was demonstrated in  \cite{Xia04}, \cite{Chen13} that for a large spreading factor we have:

\begin{equation} \label{eq_ss}
\sum\limits_{k = 1}^\beta  {\left( {x_{u,k - \tau _l } x_{u,k - \tau _j } } \right)}  \approx 0 \:\:\: for \: l \ne j \cdot
\end{equation}

Since the channel is assumed slow fading, the channel coefficients are assumed  constant during the transmission time of a MC-DCSK frame,  and change every data stream $u$. The decision variable for the $i^{th}$ bit of the $u^{th}$ data stream at the output of the correlator is:

\begin{equation}
\begin{array}{l}
 D_{u,i} \approx T_c\sum\limits_{k = 1}^\beta  {\left( {\sum\limits_{l = 1}^L {\lambda _{u,l} x_{u,k - \tau _{l} } s_{u,i} } + n_{u,k}^i } \right)}  \\ 
  \times \left( {\sum\limits_{l = 1}^L {\lambda _{u,l} x_{u,k - \tau _{l} } }+ n_{u,k}^{} } \right) \\ 
 \end{array}
 \end{equation}
\\
where  $\lambda_{u,l}$ , and  $\tau _{l}$  are  the channel coefficient and the time delay  of the $l^{th}$ path affecting the $u^{th}$ data stream respectively. The components $n_{u,k}$ and $n_{u,k}^{i}$ are two independent zero Gaussian noises coming from the reference and the $i^{th}$ bit subcarrier. For mathematical simplification we set the time chip equal to one ($Tc=1$).


 Finally, based on equation (\ref{eq_ss}), the decision variable may be approximated as:

\begin{equation}\label{dec_f}
\begin{array}{l}
 D_{u,i}  \approx  \sum\limits_{k = 1}^\beta  {\sum\limits_{l = 1}^L {\lambda _{u,l}^2 x_{_{u,k - \tau _{l - 1} } }^2 s_{u,i} } }  +  \\ 
\sum\limits_{k = 1}^\beta  {  \sum\limits_{l = 1}^L {\lambda _{u,l} x_{u,k - \tau _{l - 1} } } \left( {s_{u,i} n_{u,k}^{}  + n_{u,k}^i } \right) } + \sum\limits_{k = 1}^\beta  {\left( {n_{u,k}^i n_{u,k}^{} } \right)}  \\ 
 \end{array}
 \end{equation}
\\
 The $i^{th}$ bit of the $u^{th}$ data stream is decoded by comparing the output $D_{u,i}$ to a threshold of zero. 
 
In the decision variable given in equation (\ref{dec_f}), the first term is the useful signal, while the second and third are zero-mean additive noise interferences.  

The output of the correlator for the MC-DCSK of equation (\ref{dec_f}) can be written in the form

\begin{equation}\label{dv_ff}
D_{u,i}  = s_{u,i} \frac{{ (M-1)\sum \limits_{l = 1}^L {\lambda _{u,l}^2} E_{b}^{(u)} }}{M} + W + Z,
\end{equation}

\[E_{b}^{(u)}  = \frac{{M}}{{M-1}} \sum\limits_{k = 1}^\beta  {x_{u,k}^2 },\]
\\
where $E_{b}^{(u)}$ is the transmitted bit energy for a given data sequence $u$.

\[ W = \sum\limits_{k = 1}^\beta  {  \sum\limits_{l = 1}^L {\lambda _{u,l} x_{u,k - \tau _{l - 1} } } \left( {s_{u,i} n_{u,k}^{}  + n_{u,k}^i } \right) }, \]

\[ Z = \sum\limits_{k = 1}^\beta  {\left( {n_{u,k}^i n_{u,k}^{} } \right)} \cdot \]

For a given $i^{th}$ bit of an $u^{th}$ data stream, the instantaneous mean and variance of the decision variable are derived as follows:

\begin{equation} \label{mean}
E(D_{u,i}) = s_{u,i}\frac{{(M-1) \sum \limits_{l = 1}^L {\lambda _{u,l}^2} E_{b}^{(u)} }}{M} \cdot
\end{equation}

Since the three terms of (\ref{dv_ff}) are uncorrelated,  the noise samples and channel coefficients  are independent. The conditional variance of the decision variable for a given bit $i^{th}$ is:

\begin{equation}
\begin{array}{l}
 Var(D_{u,i} ) = E\left( {( \frac{{M-1}}{{M}} \sum \limits_{l = 1}^L {\lambda _{u,l}^2}  E_{b}^{(u)} s_{u,i} )^2 } \right) + \\ 
E\left( \hspace{-0.1cm} {(\sum\limits_{k = 1}^\beta  { \sum \limits_{l = 1}^L {\lambda _{u,l}} x_{u,k} n_{u,k  }^{i} } )^2 } \hspace{-0.1cm} \right) \hspace{-0.1cm} +\hspace{-0.1cm} E \hspace{-0.1cm} \left(\hspace{-0.1cm}  {(\sum\limits_{k = 1}^\beta  {  \sum \limits_{l = 1}^L {\lambda _{u,l}} x_{u,k}s_{u,i} n_{u,k} } )^2 } \hspace{-0.1cm} \right)  \\ 
 + E\left( {(\sum\limits_{k = 1}^\beta  {n_{u,k} n_{u,k  }^{i} } )^2 } \right)-\left(\frac{{(M-1)}}{{M}}  \sum \limits_{l = 1}^L {\lambda _{u,l}^2} E_{b}^{(u)} s_{u,i} \right)^2, \\ 
 \end{array}
\end{equation}

Finally after simplifications, 
\begin{equation}\label{var}
Var(D_i ) = \frac{{(M-1)E_{b}^{(u)} }}{M}\sum \limits_{l = 1}^L {\lambda _{u,l}^2}  N_0 /2    + \beta  N_0^2 /4 \cdot
\end{equation}

In order to compute the BER with our approach, the error probability must  be evaluated first for a given received energy $E_{b}^{(u)}$ and channel coefficient $\lambda_{u,l}$. Considering the bit energy (or chaotic chips) as a deterministic variable, the decision variable at the output of the correlator is necessarily a random Gaussian variable. Using equations (\ref{mean}) and (\ref{var}), the bit error probability is :

 \begin{equation}\label{ber_ga_ex}
\begin{array}{l}
BER  = \\
\frac{1}{2}\Pr \left( {\left. {D_{u,i}  < 0} \right|s_{u,i}  =  + 1} \right) + \frac{1}{2}\Pr \left( {\left. {D_{u,i}  > 0} \right|s_{u,i}  =  - 1} \right) \\ 
 \,\,\,\,\,\,  \,\,\,\,\, = \frac{1}{2}erfc\left( {\frac{{E\left[ {\left. {D_{u,i} } \right|s_{u,i}  =  + 1} \right]}}{{\sqrt {2{\mathop{\rm Var}} \left[ {\left. {D_{u,i} } \right|s_{u,i}  =  + 1} \right]} }}} \right), \\ 
 \end{array}
\end{equation}
\\
where $erfc(x)$ is the complementary error function defined by:

\[
erfc(x) \equiv \frac{2}{{\sqrt \pi  }}\int_x^\infty  {e^{ - \mu ^2 } } d\mu 
\]

The BER expression for the MC-DCSK system is:   

\begin{equation} \label{BER}
\begin{array}{l}
BER = \\
 \hspace{-0.1cm} \frac{1}{2}erfc\left( \hspace{-0.15cm} {\left[ \hspace{-0.1cm} {\frac{{MN_0 }}{{\left( {M - 1} \right)\sum \limits_{l = 1}^L {\lambda _{u,l}^2}  E_b^{(u)} }}\hspace{-0.1cm} +  \hspace{-0.1cm} \frac{{M^2  \beta N_0^2 }}{{2\left( {M - 1} \right)^2 (\sum \limits_{l = 1}^L {\lambda _{u,l}^2} E_b^{(u)})^2 }}} \right]^{ - \frac{1}{2}} } \hspace{-0.1cm} \right)\cdot \\
 \end{array}
\end{equation}
%


Many approaches have been considered for computing the BER of chaos-based communication systems, with the most widely used being the Gaussian approximation, which considers the transmitted bit energy $E_{b}^{(u)}$ as constant \cite{Sus00}. This assumption gives a good approximation of the performance for high spreading factors. Based on this fact, the overall BER expression of the MC-DCSK system can be simplified as:



\begin{equation} \label{dcsk_int_s}
BER =  \hspace{-0.15cm} {\frac{1}{2}erfc\left( \hspace{-0.15cm} {\left[ {\frac{M}{{\left( {M - 1} \right)\gamma _b }} + \frac{{M^2 \beta }}{{2\left( {M - 1} \right)^2 \gamma _b^2 }}} \right]^{ - \frac{1}{2}} } \right)\hspace{-0.15cm} , } 
\end{equation}
\\
where $\gamma _b =  \sum \limits_{l = 1}^L {\lambda _{u,l}^2}  Eb/N_0$

For high spreading factors the bit energy $E_b$ can be assumed to be constant \cite{Kad09ieee}. In this case, and for  $L$ independent and identically distributed (i.i.d) Rayleigh-fading channels, the PDF of the instantaneous $\gamma _b$ can be written as \cite{Pro95}:

\begin{equation}
f\left( {\gamma _b } \right) = \frac{{\gamma _b^{L - 1} }}{{\left( {L - 1} \right)!\bar \gamma _c^L }}\exp \left( { - \frac{{\gamma _b }}{{\bar \gamma _c^{} }}} \right) \equiv f\left( {\gamma _b ,\bar \gamma _c^{} ,L} \right)
\end{equation} 

where $\bar \gamma _c^{}$ is the average SNR per channel defined as:

\begin{equation}
\bar \gamma _c^{}  = \frac{{E_b }}{{N_0 }}E\left( {\lambda _j^2 } \right) = \frac{{E_b }}{{N_0 }}E\left( {\lambda _l^2 } \right),\,\,\,\,\,\,\,\,j \ne l \end{equation}

For dissimilar channels, the PDF of $\gamma _b$ can be written as \cite{Pro95}:

\begin{equation} \label{pdf_ray}
\begin{array}{l}
 f\left( {\gamma _b } \right) = \sum\limits_{l = 1}^L {\frac{{\rho _l }}{{\bar \gamma _l }}} \exp \left( { - \frac{{\gamma _b }}{{\bar \gamma _l }}} \right) \\ 
  = \sum\limits_{l = 1}^L {\rho _l f\left( {\gamma _b ,\bar \gamma _l^{} ,1} \right),}  \\ 
 \end{array}
\end{equation}
\\
where

\begin{equation}
\rho _l  = \prod\limits_{j = 1,j \ne l}^l {\frac{{\bar \gamma _l^{} }}{{\bar \gamma _l^{}  - \bar \gamma _j^{} }}} , 
\end{equation}
\\
in which $\bar \gamma _l^{}$ is the average value of $ \gamma _l^{}=\lambda_{l}^{2}E_b/N_0$, which is the instantaneous SNR on the $l^{th}$ channel.

Finally, the BER expression of the MC-DCSK system under multipath  Rayleigh fading channel is:

\begin{equation} \label{dcsk_int_s}
\begin{array}{l}
BER = \\
 \hspace{-0.15cm} \int\limits_0^{ + \infty } \hspace{-0.15cm} {\frac{1}{2}erfc\left( \hspace{-0.15cm} {\left[ {\frac{M}{{\left( {M - 1} \right)\gamma _b }} + \frac{{M^2 \beta }}{{2\left( {M - 1} \right)^2 \gamma _b^2 }}} \right]^{ - \frac{1}{2}} } \right)\hspace{-0.15cm} f\left( {\gamma _b } \right)d\gamma _b } \cdot  \\ 
 \end{array}
\end{equation}


\subsection{BER computation methodology under AWGN channel }

In this section, the performance of the MC-DCSK under an AWGN channel will be evaluated for low and high spreading factors. The aim of this analysis is to highlight the non constant bit energy problem when the spreading factor is very low. In this case, one path is considered $L=1$ within a channel coefficient  equal to one $ \lambda=1$ and $\gamma _b = Eb/N_0$. 

For high spreading factors, the transmitted bit energy $E_b$ can be considered  constant. The BER expression of the MC-DCSK system may then be approximated by:

\begin{equation} \label{dcsk_int_ex}
\begin{array}{l}
BER =  \frac{1}{2}erfc\left( \hspace{-0.15cm} {\left[ {\frac{M N_0}{{\left( {M - 1} \right)E_b }} + \frac{{M^2 N_0^2 \beta }}{{2\left( {M - 1} \right)^2 E _b^2 }}} \right]^{ - \frac{1}{2}} } \right)\cdot  \\ 
 \end{array}
\end{equation}

\begin{figure}[htb!]
\centering
\includegraphics[width=8 cm]{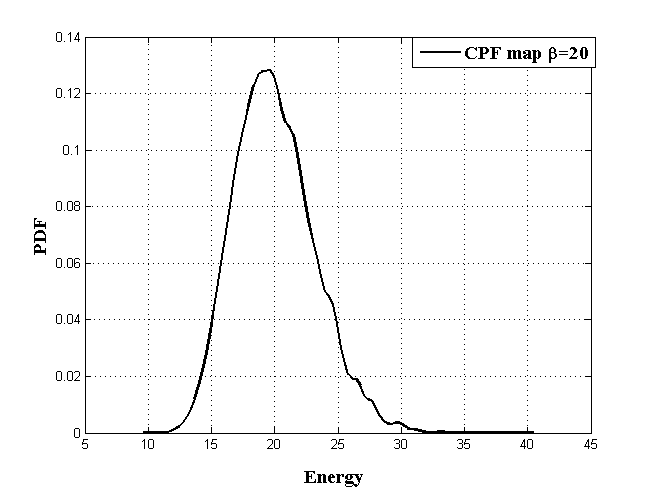}
\caption {Histogram of the bit energy distribution  $E_{b}$ for $\beta=20$ \label{pdf_cpf}}

\end{figure}

For low spreading factors, the bit energy cannot be assumed constant. In fact, because of the non-periodic nature of chaotic signals, the transmitted bit energy after spreading by chaotic sequences definitely varies from one bit to another \cite{Kad09ieee} for low spreading factors. To compute (\ref{dcsk_int_s}) in this special case, it is necessary to get the bit energy distribution for the given chaotic map. With this aim, we fitted the histogram of the energy distribution for the CPF sequence. Fig. \ref{pdf_cpf} shows the histogram of the bit energy after spreading by the CPF chaotic sequence for $\beta=20$. This histogram has been obtained using ten million samples. From these samples, energies of successive bits are calculated for a given spreading factor. The bit energy is assumed to be the output of a stationary random process \cite{Isa97}; hence the histogram obtained in Fig. \ref{pdf_cpf} can be considered as a good estimation of the probability density function of the bit energy. The BER expression of the MC-DCSK system for low spreading factors is:  

\begin{equation} \label{dcsk_int_en}
\begin{array}{l}
BER = \\
 \hspace{-0.15cm} \int\limits_0^{ + \infty } \hspace{-0.15cm} {\frac{1}{2}erfc\left( \hspace{-0.15cm} {\left[ {\frac{M}{{\left( {M - 1} \right)\gamma _b }} + \frac{{M^2 \beta }}{{2\left( {M - 1} \right)^2 \gamma _b^2 }}} \right]^{ - \frac{1}{2}} } \right)\hspace{-0.15cm} f\left( {\gamma _b } \right)d\gamma _b } \cdot  \\ 
 \end{array}
\end{equation}
\\
where 
$\gamma _b = Eb/N_0$

Given the shape of this energy distribution, the analytical expression appears difficult to compute, leaving numerical integration as a solution for performing the BER computation. The expression (\ref{dcsk_int_en}) can be computed numerically, taking into account the bit-energy variation shown in Fig. \ref{pdf_cpf} .

\subsection{Numerical integration method}

The numerical integration is performed by using the analytically PDF  given in equation (\ref{pdf_ray}) for  expressions (\ref{dcsk_int_s}), and  the histogram plotted in  figure \ref{pdf_cpf} for expression (\ref{dcsk_int_en}). Then, we can compute the BER  integral by using the following expression:

\begin{equation}\label{num_int}
\begin{array}{l}
BER \approx \\
  \sum\limits_{n = 1}^C  {\frac{1}{2}erfc\left( \hspace{-0.15cm} {\left[ {\frac{M}{{\left( {M - 1} \right)\gamma _{b,n} }} + \frac{{M^2 \beta }}{{2\left( {M - 1} \right)^2 \gamma _{b,n}^2 }}} \right]^{ - \frac{1}{2}} } \right)\hspace{-0.15cm} f\left( {\gamma _{b,n} } \right)} 
 \end{array}
\end{equation}
\\
where $C$ is  the number of histogram classes and $f\left( {\gamma _{b,n} } \right)$ is the probability of having the energy in intervals centered on $\gamma _{b,n}$. In our paper set the number of classes to  $C=100$ with a unit integration step size. 

%

%
%
%
%
%
%

\section{Simulation results and discussions}
\label{sec:Simulation}

\subsection{Performance evaluation}
To evaluate the effect of the number of subcarriers on performance, we plot the computed BER expressions with simulation results of the MC-DCSK system over  AWGN and multipath Rayleigh fading channels. The results obtained are for different numbers of subcarriers $M$ and spreading factors $\beta$.

The parameters of the simulation in a mono-user case are set as follows: the MC-DCSK system uses the square-root-raised-cosine chip waveform with a roll-off factor $\alpha$ equal to $0.25$. As shown in equation (\ref{sr_fc}), the spreading factor is computed as a function of the number of subcarriers $M$, the bit duration $T_b$, and the total allocated bandwidth $B$. In our simulations, we set the bit duration $T_b=400$, $B=1$. For $M=64$ the allowed spreading factor  $\beta =5$, for $M=16$ subcarriers $\beta =20$, for $M=8$ $\beta =40$, and for $M=2$ $\beta =160$.

\begin{figure}[htb!]
\centering
\includegraphics[width=8 cm]{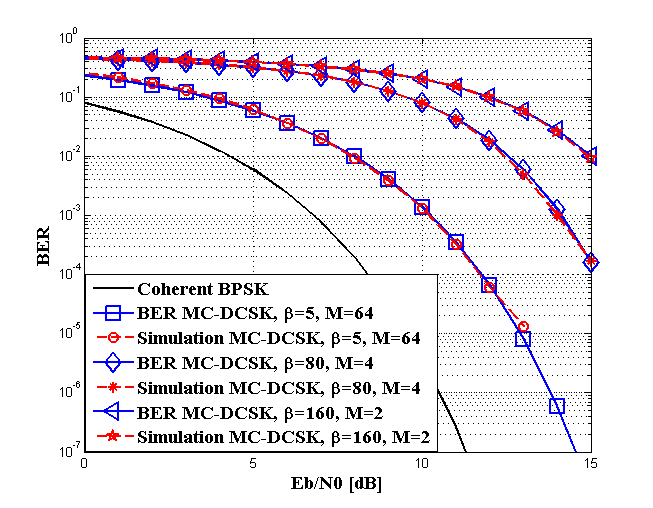}
\caption{Simulation and BER expression, for different spreading factor $\beta$ values, number of subcarrier $M$,  and under an AWGN channel \label{mc_DCSK_sf}}

\end{figure}

Fig. \ref{mc_DCSK_sf} presents the performances obtained from the BER  expression of (\ref{dcsk_int_ex}) for high spreading factors and equation (\ref{dcsk_int_en}) for  a low spreading factor ($\beta=5$).  The BER performances are plotted with the Monte Carlo simulations of the MC-DCSK system under an AWGN channel. It clearly appears that there is an excellent match between simulations and our computed BER expressions  for any number of subcarriers and spreading factors. 

\begin{figure}[htb!]
\centering
\includegraphics[width=8 cm]{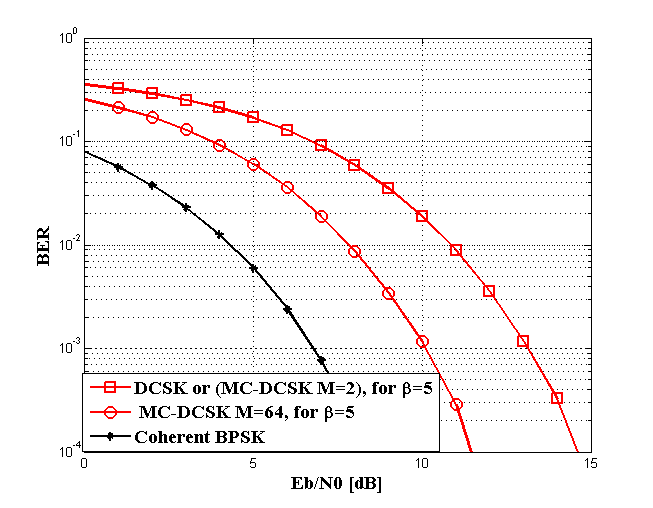}
\caption{BER comparison of MC-DCSK for M=64 and DCSK (i.e MC-DCSK for M=2) where the spreading factor $\beta=5$ under AWGN channel \label{DCSK_mc_DCSK}}

\end{figure}

In Fig. \ref{DCSK_mc_DCSK} we study the effect of the number of subcarriers  on the system performance under an AWGN channel. To that end, we set the spreading factor to $\beta=5$ and the bit duration $T_b$, and then we assume that the bandwidth $B$ is wide enough to support any number of subcarriers $M$. Fig. \ref{DCSK_mc_DCSK} shows  interesting results of our proposed MC-DCSK system in terms of performance enhancement. In fact, for a given spreading factor, when the number of subcarriers $M$ increases, the $DBR$ ratio tends toward one, meaning that less reference energy is used to transmit one bit. In other words, the reference energy is shared among $M-1$ bits. This performance improvement, proven in the BER expression, means that for a high number of subcarriers $M$, we need less energy to reach a given BER. In the same figure, we show the performance improvement by simulation for $M=2$ and $M=64$, with a fixed spreading factor equal to $\beta=5$. In the case of  $M=2$, the MC-DCSK system is equivalent to a DCSK system. The results shown in Fig. \ref{DCSK_mc_DCSK} can be seen as a  performance comparison  between the proposed system with that of the conventional DCSK. In the same figure, we can observe a degradation in performance between the MC-DCSK system for $M=64$ and the coherent BPSK one. This degradation comes from the two noise sources added to the reference and data carrier signals.

\begin{figure}[htb!]
\centering
\includegraphics[width=8 cm]{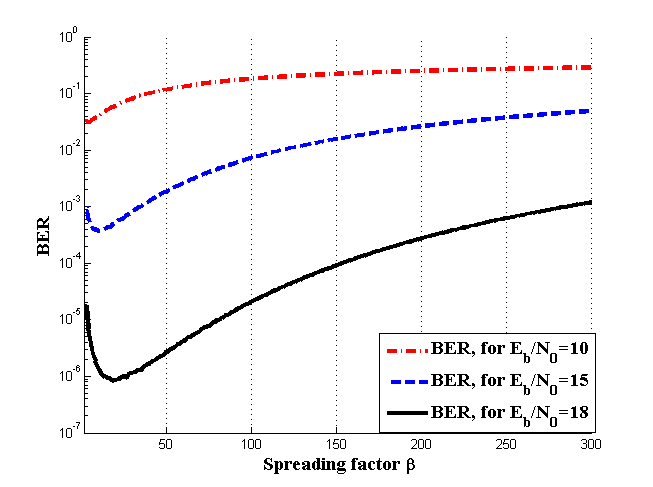}
\caption{BER versus the spreading factor $\beta$ for MC-DCSK for M=2 \label{DCSK_mc_sf}}
\end{figure}

To understand the performance behavior of a MC-DCSK system for different spreading factors the optimal spreading factor must be discussed. The optimal spreading factor was studied in \cite{yao05} for single carrier a DCSK system. Fig.\ref{DCSK_mc_sf} evaluates the effect of the value of the spreading factor on the performance of the MC-DCSK under  an AWGN channel. The simulated bit error rate is plotted for different values of the spreading factor $\beta$ with a fixed $E_b/N_0$ and a number of subcarriers $M=2$. The bandwidth is assumed to be wide enough to support any spreading factor value. Because the BER expression is approximated and computed by numerical integration, the theory in this case can only qualitatively describe the dependence of the MC-DCSK spreading factor. Simulation  shows that on the spreading factor values between $ 5$ and $50$ minimize bit error rates at fixed $E_b/N_0$. From  these results, we see that good performances are obtained for low spreading factor values which makes this system implementation feasible even for a moderate bandwidth.

\begin{figure}[htb!]
\centering
\includegraphics[width=9 cm]{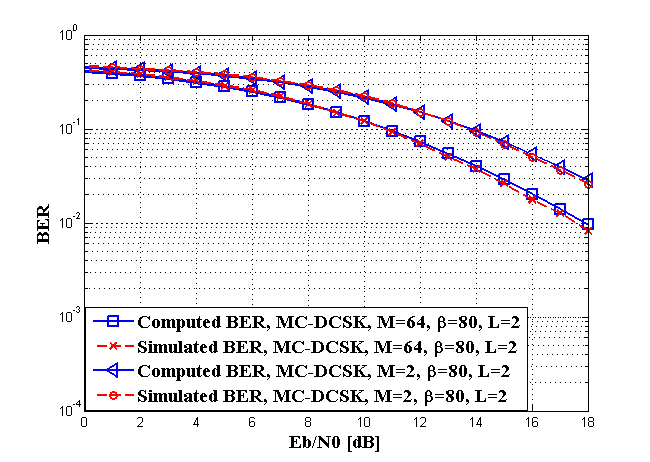}
\caption{Simulation and BER expression, for a spreading factor  $\beta=80$,  number of subcarriers  $M=2$, $M=64$ under multipath Rayleigh fading  channel $L=2$ with equal average  power gain  $E(\lambda_1^2)= E(\lambda_2^2)= \frac{1}{2}$ and $\tau_1=0$ and $\tau_2=2$.   \label{BER_L_EG_80_64_2_L_2}}

\end{figure}

\begin{figure}[htb!]
\centering
\includegraphics[width=9 cm]{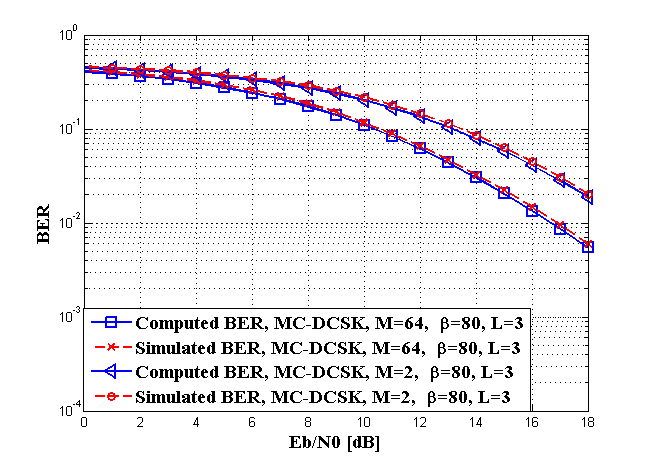}
\caption{Simulation and BER expression, for a spreading factor  $\beta=80$,  number of subcarriers $M=2$ $M=64$ under multipath Rayleigh fading channel $L=3$ with average  power gains  $E(\lambda_1^2)=4/7$, $E(\lambda_2^2)=2/7$, $E(\lambda_3^2)=1/7$ with $\tau_1=0$, $\tau_2=3$, and $\tau_3=6$.    \label{BER_L_NEG_80_64_2_L_3}}
\end{figure}

Figs. \ref{BER_L_EG_80_64_2_L_2} and \ref{BER_L_NEG_80_64_2_L_3} evaluate the effect of the multipath Rayleigh channel on the performance of the MC-DCSK system. The  bit error rate expression given in equation (\ref{dcsk_int_s}) is plotted with the computer simulation. The system's performance plotted in Fig.\ref{BER_L_EG_80_64_2_L_2} is evaluated for two different subcarriers $M=2$ and $M=64$, a spreading factor equal to $ \beta=80$, and for two paths $L=2$ having an equal average power gain  $E(\lambda_1^2)=E(\lambda_2^2)= \frac{1}{2}$ with $\tau_1=0$ and $\tau_2=2$.

In Fig.\ref{BER_L_NEG_80_64_2_L_3}, the performance is evaluated for two different subcarriers $M=2$ and $M=64$, a spreading factor equal to $ \beta=80$. In this case,  three paths $L=3$ are considered with different average  power gains. The average  power gain of the third path is $3$ dB below the second path and the average power gain of the second path is $3$ dB below the line-of-sight path with the appropriate time delays $\tau_1=0$, $\tau_2=3$, and $\tau_3=6$. 

It clearly appears that there is an excellent match between simulations and our computed BER expressions for any number of subcarrier, number of path, and average power gain. The results shown in Figs. \ref{BER_L_EG_80_64_2_L_2} and \ref{BER_L_NEG_80_64_2_L_3} confirm the exactitude our of assumption.

\begin{figure}[htb!]
\centering
\includegraphics[width=9 cm]{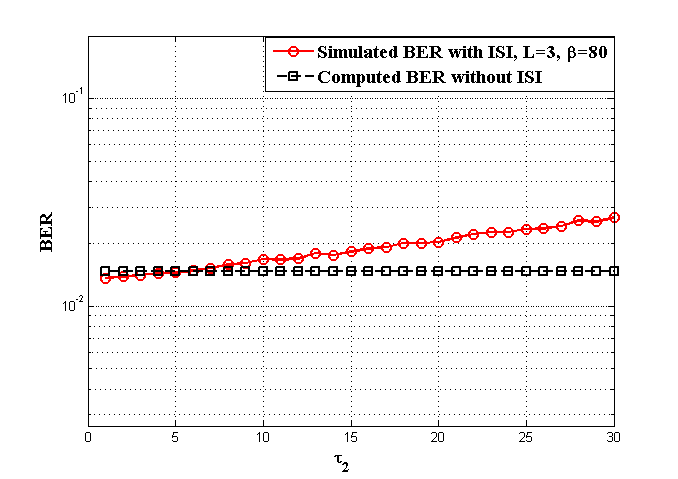}
\caption{BER perfomance of MC-DCSK sysytem over 3 paths Rayleigh fading channel  with  average gain powers  $E(\lambda_1^2)=4/7$, $E(\lambda_2^2)=2/7$, $E(\lambda_3^2)=1/7$ versus $\tau_2$ and $\tau_3=\tau_2+1$, for a spreading factor  $\beta=80$, number of subcarriers $M=64$.    \label{limit_L_3_beta_80}}
\end{figure}

Fig.\ref{limit_L_3_beta_80} shows the effect of the time delay on the BER performance. The results are obtained for  a fixed $Eb/N_0=15$ dB, a spreading factor equal to $\beta=80$, and number of subcarriers $M=64$. Three paths Rayleigh fading channels are considered with  average gain powers $E(\lambda_1^2)=4/7$, $E(\lambda_2^2)=2/7$, $E(\lambda_3^2)=1/7$. The performance are plotted in this figure for different values of time delays $\tau_2$ for the second path and $\tau_3=\tau_2+1$ for the third path. The simulation results shows that the ISI can be neglected when the time delays are much less than the bit duration.  In addition, Fig.\ref{limit_L_3_beta_80} shows  that the limit of the  negligible ISI assumption is still valid up until time delays  $\tau_2=12$ and $\tau_3=13$. On the other hand,  when the time delay is large, the computed performance does not agree with the simulated one, because in the derivation of the BER expression, the time delay is assumed small compared to the bit duration and hence the ISI was neglected. 

\vspace{+0.7cm}

\subsection{Discussions}

The proposed system meets the following properties:

\begin{itemize}
\item Non-coherent system: Robust receiver;
\end{itemize}

\begin{itemize}
\item Spread spectrum system: resistance to interference;  
\end{itemize}

\begin{itemize}
\item Chaotic signals: easy to generate, low PAPR in multi-carrier transmissions and good correlation properties; 
\end{itemize}

\begin{itemize}
\item Multi-carrier DCSK: high spectral efficiency and low power consumption.
\end{itemize}

\section{Conclusion}
\label{sec:Conclusion}
An energy-efficient non-coherent multi-carrier spread spectrum system has been presented. From the outstanding energy inefficiency drawback imposed by time-multiplexed differential modulations, a novel frequency multiplexed architecture is designed. The multi-carrier characteristic of this novel design enables significant energy savings and a higher spectral efficiency as compared to differential systems because in the new system, the reference signal is only sent once for many parallel bits. The energy efficiency of the proposed system is analyzed and a $DBR$ is derived, with results showing that for $M > 20$ subcarriers, the energy lost in transmitting the reference is less than $ 5\%$ of the total bit energy per bit. The performance of the proposed system is studied, and  the bit error rate expressions are derived for an AWGN and  multipath  Rayleigh fading channels. Simulation results match the theoretical BER expressions, justifying our approximations and demonstrating the accuracy of our approach. To compare the performance of the proposed system with that of the DCSK, the simulated BERs are plotted with the same spreading factor, where results show an increase in performance as compared to the conventional DCSK.  Our future work will focus on defining multi-user access strategies and performance improvements of this system.

\bibliographystyle{IEEEtran}
\bibliography{bibliographie_twir}

%

%
%

\begin{IEEEbiography}[{\includegraphics[width=1in,height=1.25in,clip,keepaspectratio]{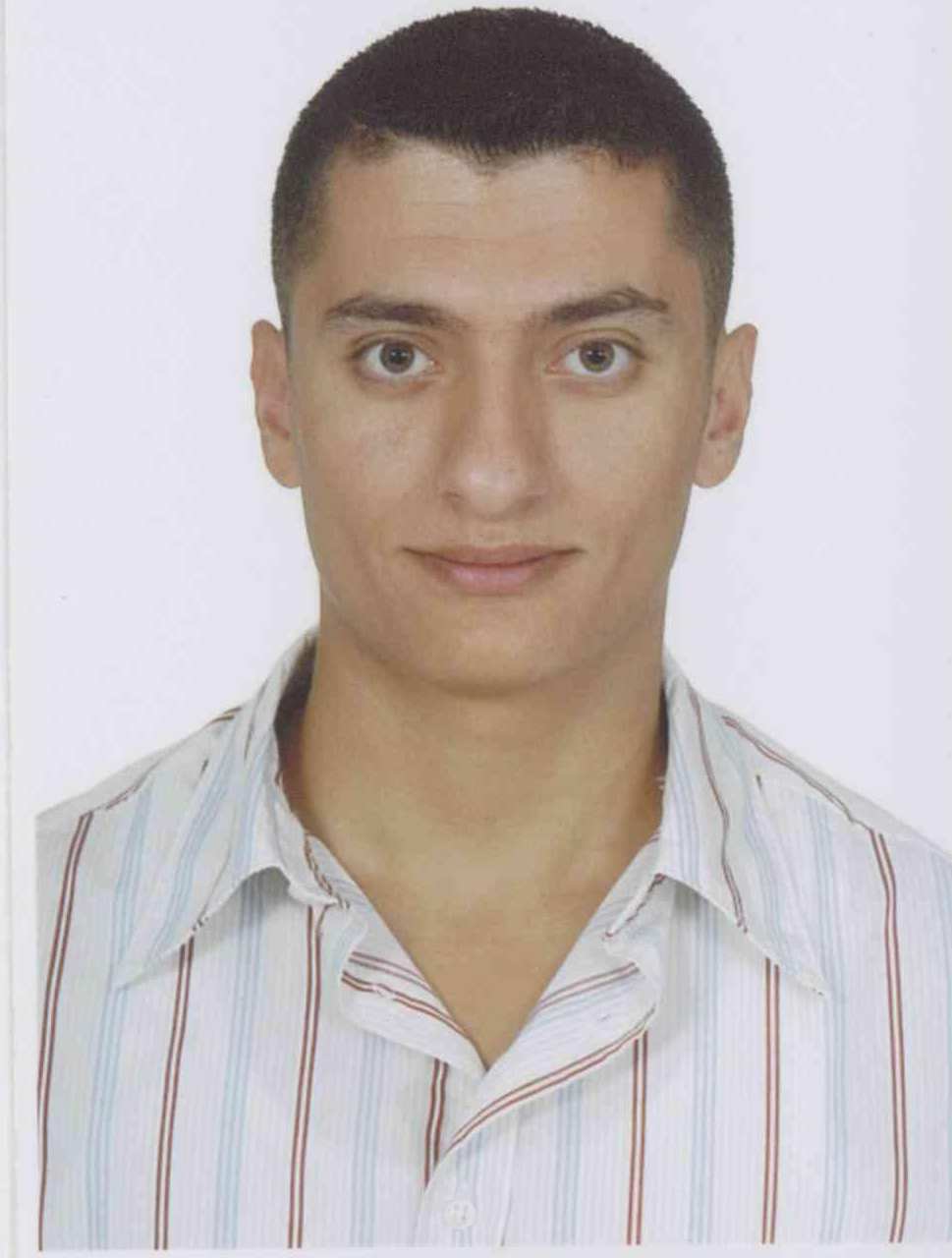}}] {Georges Kaddoum}
earned a bachelor's degree from the Lebanese University, Beyrouth, Lebanon with a First Class Honours Degree in physics \& electronics in 2003, the B. Eng. from \'{E}cole Nationale sup\'{e}rieure des techniques avanc\'{e}es (ENSTA), and the M.Sc. degree in circuits, systems, and signal processing from Telecom Bretagne, Brest, France both in 2005. He received the Ph.D. degree in Telecommunications with distinction from University of Toulouse, Toulouse, France in 2008. In 2008, he was a lecturer in digital communications, and signal processing at the Institut National Polytechnique de Toulouse (University of Toulouse, INP-ENSEEIHT). From 2009 to 2011, he was Postdoctoral Research Fellow with the Department of Electrical Engineering, University of Quebec, \'{E}cole de technologie supérieure (\'{E}TS), Montreal, Quebec, Canada. Since 2008, Georges Kaddoum has been with the \'{E}TS where he is currently a scientific researcher. 
His recent research activities cover wireless communication systems, chaotic modulations, secure transmissions, and space communications \& Navigation. He published over 45 journal and conference papers to date and holds one pending patent. He is currently an editorial board of the CIP Journal Wireless Communications and Networking.

\end{IEEEbiography}

\begin{IEEEbiography}[{\includegraphics[width=1in,height=1.25in,clip,keepaspectratio]{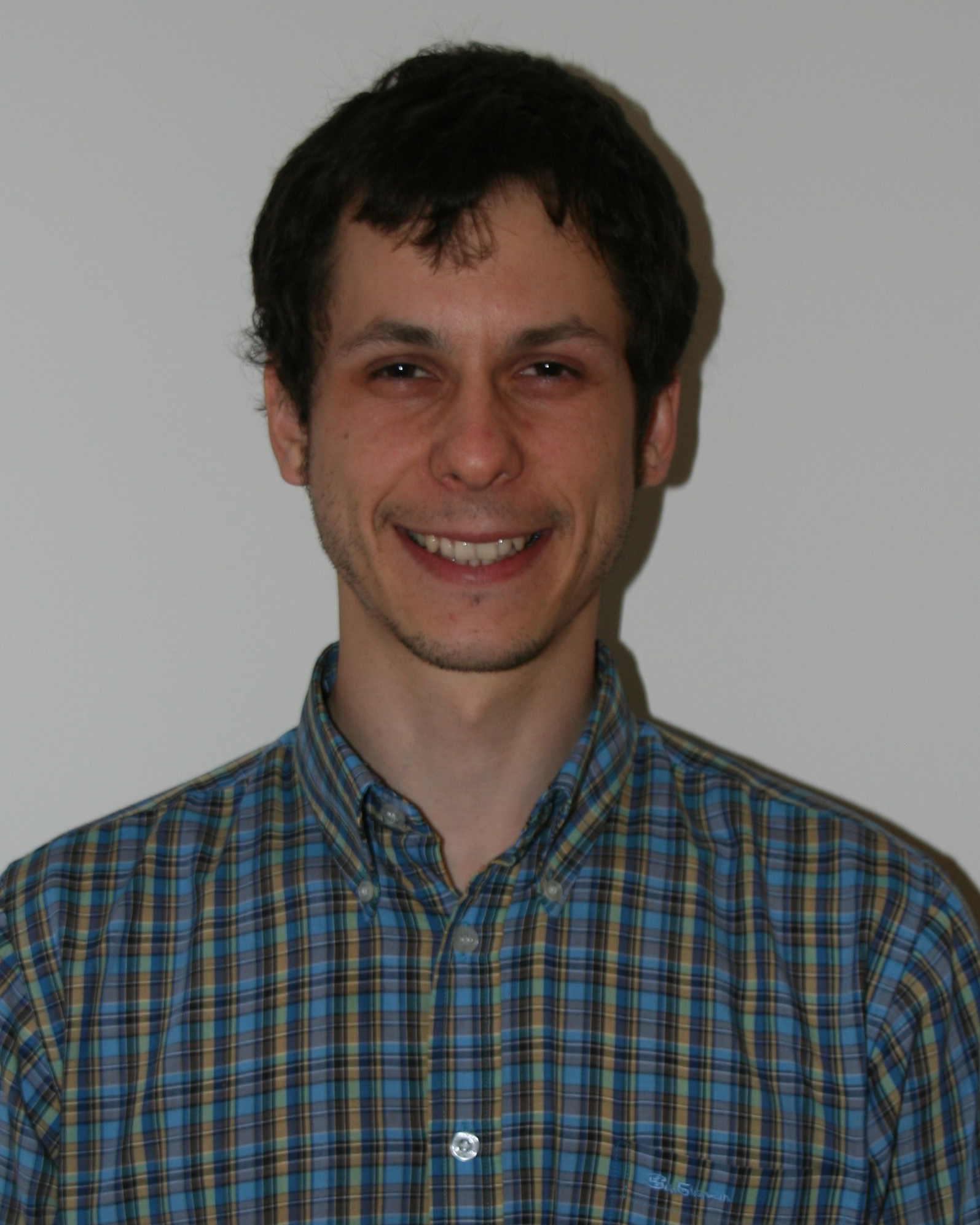}}] {Fran\c cois-Dominique Richardson}
 received the B.Eng. and M.Eng degrees in electrical engineering from University of Quebec, \'{E}cole de technologie supérieure (\'{E}TS), Montreal, Quebec, Canada, in 2009 and 2011 respectively. From 2011 to 2013, he was a research fellow with the NSERC Ultra Electronics Chair, Wireless Emergency and Tactical Communication, \'{E}TS. His research interest covers low-power systems, adaptive systems and wireless communications. He is  actually with Octasic as an ASIC designer engineer. 
\end{IEEEbiography}

\begin{IEEEbiography}[{\includegraphics[width=1in,height=1.25in,clip,keepaspectratio]{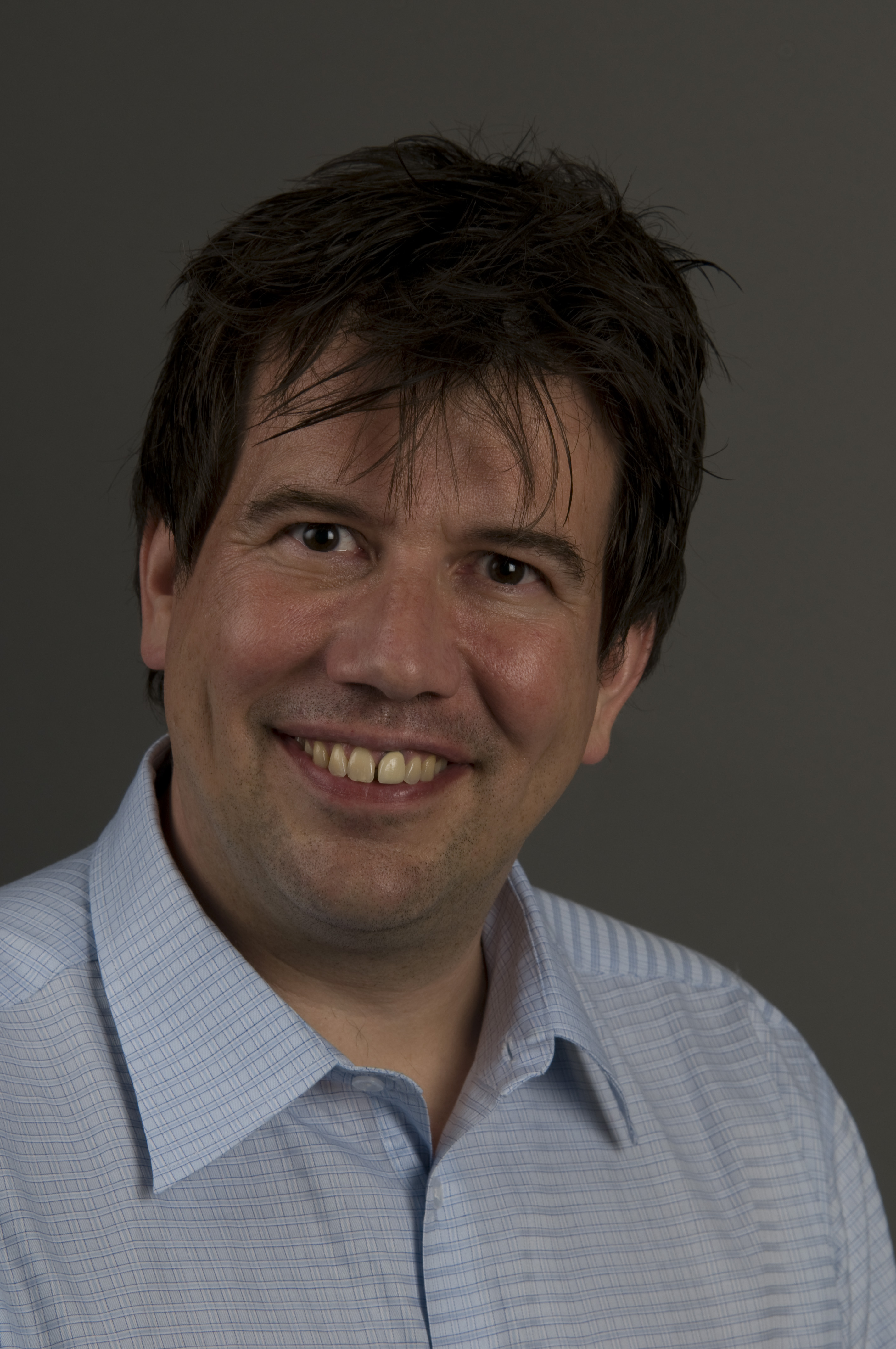}}]%
{ Fran\c cois Gagnon}
 received the B.Eng. and Ph.D. degrees in electrical engineering from \'{E}cole Polytechnique de Montreal, Montreal, Quebec, Canada. Since 1991, he has been a Professor with the Department of Electrical Engineering, École de Technologie Supérieure, Montreal, Quebec, Canada. He chaired the department from 1999 to 2001, and is now the holder of the NSERC Ultra Electronics Chair, Wireless Emergency and Tactical Communication, at the same university. His research interest covers wireless high-speed communications, modulation, coding, high-speed DSP implementations, and military pointto- point communications. He has been very involved in the creation of the new generation of high-capacity line of-sight military radios offered by the Canadian Marconi Corporation, which is now Ultra Electronics Tactical Communication Systems. The company has received, for this product, a "Coin of Excellence" from the U.S. Army for performance and reliability. Prof. Gagnon was awarded the 2008 NSERC Synergy Award for the fruitful and long lasting collaboration with Ultra Electronics TCS.
\end{IEEEbiography}

\end{document}